\documentclass[prb,twocolumn,superscriptaddress,floatfix,letterpaper,url]{revtex4-1}

\newcommand{\nn}{\nonumber}
\newcommand{\vsc}{v_{\text{sc}}}

\begin{document}

\title{
A mechanism of $\frac12 \frac{e^2}{h}$ conductance plateau without 
1D chiral Majorana fermions
}

\author{Wenjie Ji}
\affiliation{Department of Physics, Massachusetts Institute of
Technology, Cambridge, Massachusetts 02139, USA}

\author{Xiao-Gang Wen}
\affiliation{Department of Physics, Massachusetts Institute of
Technology, Cambridge, Massachusetts 02139, USA}

\begin{abstract} 
We address the question about the origin of the $\frac12 \frac{e^2}{h}$
conductance plateau observed in a recent experiment on an integer quantum Hall
(IQH) film covered by a superconducting (SC) film. Since 1-dimensional (1D)
chiral Majorana fermions on the edge of the above device can give rise to the
half quantized plateau, such a plateau was regarded as a smoking-gun evidence
for the chiral Majorana fermions.  However, in this paper we give another
mechanism for the $\frac12 \frac{e^2}{h}$ conductance plateau. We find the
$\frac12 \frac{e^2}{h}$ conductance plateau to be a general feature of a good
electric contact between the IQH film and SC film, and cannot distinguish the
existence or the non-existence of 1D chiral Majorana fermions.  We also find
that the contact conductance between SC and an IQH edge channel has a non-Ohmic
form $\sigma_\text{SC-Hall} \propto V^2$ in $k_BT \ll eV$ limit, if the SC and
IQH bulks are fully gapped.

\end{abstract}

\maketitle

\noindent
\textbf{Introduction}:
The Majorana fermion, that is also its own anti-particle, has attracted a lot
of attention recently due to several mixed-up reasons.  One reason is the
topological quantum computation \cite{K032}, which can be realized using
non-abelian topological orders  that contain Ising non-abelian anyons, or
other more general non-abelian anyons \cite{W9102,MR9162}.  Although Ising
non-abelian anyons cannot perform universal topological quantum computation
\cite{FLZ0205}, they can be realized by non-interacting fermion
systems, such as the vortex in $p+\ii p$ 2D superconductors
\cite{RG0067,I0168,FK07071692}.  The vortex is a non-abelian anyon since it
carries a Majorana zero-mode.  Unfortunately, the zero-mode (which is not even
a particle, not to mention a fermion) was regarded as Majorana fermion, and the
search for non-abelian anyon becomes the search for Majorana fermion
\cite{qpart2016,W161003911}. Confusing statements were made, such as ``Majorana
fermions carry non-abelian statistics'' (instead of Fermi statistics).

Another reason is that Majorana fermion, proposed by Majorana in 1937 as a
possible 3D elementary particle, has not been found among elementary particles.
It will be really nice to realize the Majorana fermion in condensed matter
systems.  However, 3D Majorana fermion, defined as fermion with \emph{only}
fermion-number-parity conservation, has long been realized in superconductor
(with spin-orbital coupling) \cite{qpart2016,W161003911}.  Such a particle was
called Bogoliubov quasiparticle.  But, many do not regard Bogoliubov
fermion as Majorana fermion, and the quest to find Majorana fermion continues.

Recently, \Ref{HW160605712} claimed to discover 1D chiral Majorana fermion.
Such a discovery is new since the 1D chiral Majorana fermion is not the 3D
Majorana fermion proposed by Majorana.  1D chiral Majorana fermions are
fermions with only fermion-number-parity conservation that propagate only in
one direction in 1D space.  In 1993 \cite{W9355}, such 1D chiral Majorana
fermions were predicted to exist on the edge of some non-abelian fractional
quantum Hall states \cite{W9102,MR9162}.  In fact, the appearance of an odd
number of 1D chiral Majorana fermion modes on the edge implies the appearance
of non-abelian anyon in the bulk \cite{WZ9290,W9355}.  The non-abelian states
may have already been realized in experiments
\cite{WES8776,DM08020930,RMM0899}.  In particular, the recently observed  half
quantized thermal Hall conductance \cite{BS171000492} from the quantum Hall
edge states \cite{W9038,W9355,KF9732} provides a smoking-gun evidence of 1D
chiral Majorana fermions and its parent non-abelian fractional quantum Hall
states.  

In 2000 \cite{RG0067}, 1D chiral Majorana fermions were predicted to exist on
the edge of $p+\ii p$ 2D superconductors.  More recently, 1D chiral Majorana
fermions were found to exist on the interface of ferromagnet and superconductor
on the surface of topological insulator \cite{FK07071692}, and on the edge of
an $\nu=1$ IQH film covered by a SC film \cite{CZ10082003,WZ150700788}.

In \Ref{CZ10082003,WZ150700788}, it was shown that 1D chiral Majorana fermions
can give rise to $\frac12 \frac{e^2}{h}$ conductance plateau for a two terminal
conductance $\si_{12}$ across a Hall bar covered by a superconducting film.  In
\Ref{HW160605712}, such $\frac12 \frac{e^2}{h}$ conductance plateau was
observed in an experiment on stacked IQH film and SC film, which was regarded
as a ``distinct signature'' of 1D chiral Majorana fermions.  The discovered
Majorana fermions were named ``angel particles'' \cite{dailymail2017},
and have attracted a lot of attention.
However, in this paper, we will show  that the $\frac12 \frac{e^2}{h}$
conductance plateau does not imply the existence (nor the non-existence) of 1D
chiral Majorana fermions. More experiments are needed, such as the thermal Hall
experiment \cite{KF9732}, to reveal the existence of  1D chiral Majorana
fermions.

Logically speaking, even though 1D chiral Majorana fermions can give rise to
$\frac12 \frac{e^2}{h}$ conductance plateau, there are other scenarios without
chiral Majorana fermions in which $\frac12 \frac{e^2}{h}$ conductance plateau
can appear.  For example, in Fig. 4A of the very same paper \cite{HW160605712},
$\frac12 \frac{e^2}{h}$ conductance was observed in a stacked IQH film and a
metal film without the Majorana fermions. 
Similarly, \Ref{CL160800237} pointed out that $\frac12 \frac{e^2}{h}$
conductance can appear when the Hall bar under the SC film is in a metallic
state without the Majorana fermions. 

Such an explanation was discarded in \Ref{CL160800237} since it was thought to
be inconsistent with the observed magnetic field $B$ dependence of
$\sigma_{12}$ (Fig. 2C and Fig. 4A in \Ref{HW160605712}). In the experiment,
$\sigma_{12} (B)$ is found to be $\frac12 \frac{e^2}{h}$ at hight field $B$
where the topped film is normal metallic state. Then it increases up to $
\frac{e^2}{h}$, as $B$ is reduced and the topped film becomes SC.  As $B$ is
reduced further, $\sigma_{12}$ drops to a $\frac12 \frac{e^2}{h}$ plateau near
$B_c$, and then to near $0$.

\noindent \textbf{Result}: In this paper, we study the mechanism of
\Ref{CL160800237} for the $\frac12 \frac{e^2}{h}$ conductance plateau in 
detail.  We find that the Majorana-fermionless mechanism can explain the
observed curve of conductance $\sigma_{12}(B)$ very well.  The $\frac12
\frac{e^2}{h}$ conductance plateau can be a general feature of a good electric
contact between the IQH and the SC films, regardless if the 1D chiral Majorana
fermions exist or not.

\begin{figure}[tb] 
\centering \includegraphics[scale=0.36]{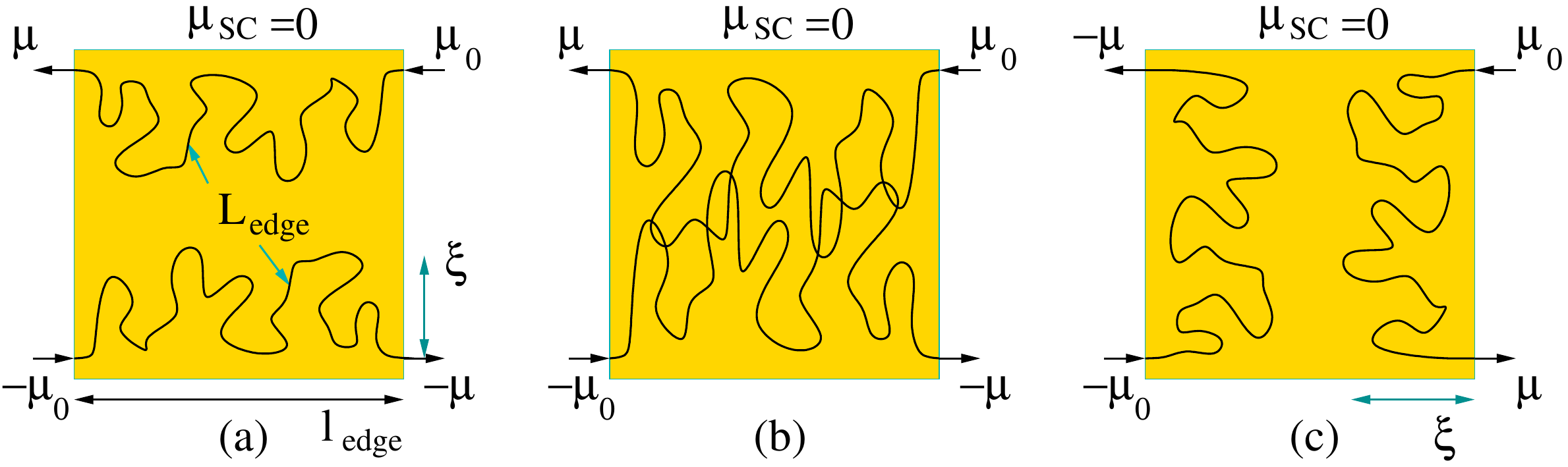} 
\caption{
A Hall bar covered by a SC film.  The Hall bar under the
superconductor can be in (a) a Chern number $N_\text{Chern}=1$ IQH phase
($B>B_c$), (b) a metallic phase, and (c) a Chern number $N_\text{Chern}=0$
insulating phase ($B<B_c$), depending on the correlation length $\xi$ of the
percolation model.
}
\label{HallSC} 
\end{figure}

\noindent \textbf{A general understanding for two terminal conductance
$\sigma_{12}$:} 
In the experiment \cite{HW160605712}, the SC layer is directly
deposited on the Hall bar.  Naively, one would expect the contact resistance,
$1/\si_\text{SC-Hall}$, between the superconductor and the edge channels of the
Hall bar under the superconductor, to be much less than
$\frac{h}{e^2}=25812\Om$.  In this case, the  two terminal conductance
$\si_{12} = \frac12 \frac{e^2}{h}$. To see this,
we assume the superconductor to have a vanishing chemical
potential $\mu_{SC}=0$ and there is no net current flowing in or out of the
superconductor.  So the chemical potentials on the two incoming edge channels
of the Hall bar should be opposite: $\mu_0$ and $-\mu_0$.  The chemical
potentials on the two outgoing edge channels of the Hall bar are also opposite:
$\mu$ and $-\mu$ (see Fig.  \ref{HallSC}).

When the contact resistance $1/\si_\text{SC-Hall}$ is low, the chemical
potentials on the two outgoing edge channels vanish: $\mu=\mu_{SC}=0$, and the
two terminal conductance $\si_{12}$ is given by $ \si_{12}
=\frac{\mu_0-(-\mu)}{\mu_0-(-\mu_0)}=\frac12$.  (In this paper, all conductance
are measured in unit of $\frac{e^2}{h}$.) We see that the $\frac12$ quantized
conductance of $\si_{12}$ is a very general feature of good contact between the
superconductor and the Hall bar under the superconductor, and one might expect
that the two terminal conductance $\si_{12}$ to be always $\frac12$.

But in the experiment, $\si_{12}\approx 1$ is observed for certain range of
magnetic field.  If we assume the superconductor and the Hall bar decouples
electronically, $\si_{12}$ should be $1$, as contributed purely from the IQH
bar.  Thus the observed $\si_{12}\approx 1$ implies that the contact resistance
between the superconductor and the Hall bar can be much larger than
$\frac{h}{e^2}$ (as observed directly via the measurement of $\si_{13}$ shown
in Fig.4C in \Ref{HW160605712}).

The observed $\sigma_{12}=\frac12$ at high field, where the topped film is
metallic, indicates the contact resistance between the metal film and the Hall
bar is always much less than $\frac{h}{e^2}$.  But in the low field region
where the film above IQH layer becomes SC, the measured $\sigma_{12}$ varies
from $1$ to $\frac{1}{2}$ depending on $B$, indicating that the contact
resistance $1/\sigma_{\text{SC-Hall}}$ between the SC film and the Hall bar can
become much bigger than $\frac{h}{e^2}$, as well as much smaller.  In this
paper, we explain such a striking pattern of the contact conductance
$\si_\text{SC-Hall}$ via a percolation model.

As the magnetic field $B$ is reduced through the critical value $B_c$, the Hall
bar under the superconductor changes from a Chern number $N_\text{Chern}=1$ IQH
state to a Chern number $N_\text{Chern}=0$ insulating state.  We use a
percolation model to describe such a transition.  In the percolation model,
when $B$ is reduced through $B_c$, the chiral edge channels of IQH state become
more and more wiggled.  Correspondingly, the Hall bar under the superconductor
has three phases: the $N_\text{Chern}=1$ phase in Fig.  \ref{HallSC}a and the
$N_\text{Chern}=0$ phase in Fig. \ref{HallSC}c, where the IQH edge channel can
be straight and short if $B$ is far away from $B_c$. Thus the contact
resistance $1/\si_\text{SC-Hall}$ is high.  The third phase is a metallic phase
in Fig.  \ref{HallSC}b, where the IQH edge channel fills the sample and is
long. As a result, the contact resistance  $1/\si_\text{SC-Hall}$ is low.

\begin{figure}[tb] 
\centering 
\includegraphics[scale=0.37]{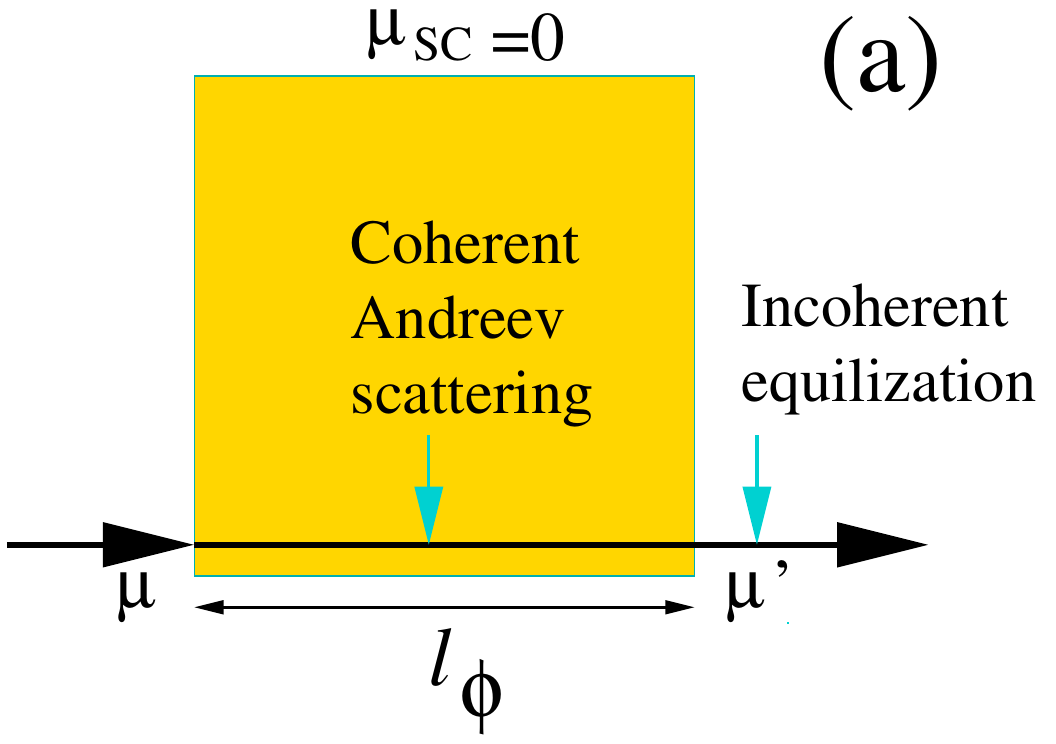} 
\hfill 
\includegraphics[scale=0.3]{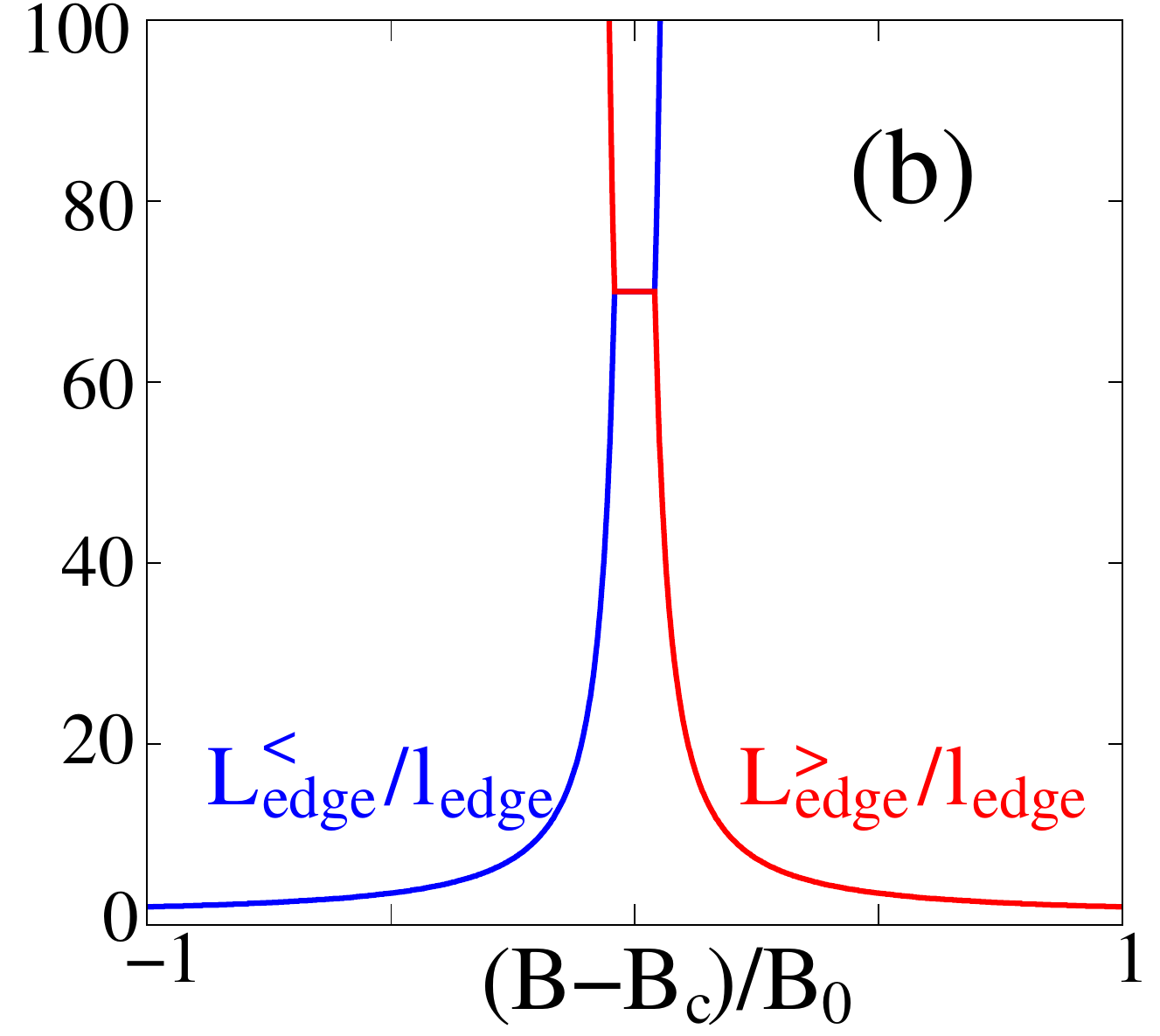}
\caption{
(a) A segment of IQH edge under a superconductor.
(b) $L_\text{edge}^>$ and $L_\text{edge}^<$ as a function of $B$.
}
\label{SCedge} 
\end{figure}

\noindent \textbf{A microscopic calculation of the contact conductance
$\si_\text{SC-Hall}$ between the superconductor and an IQH edge channel}: We
first assume the SC film and IQH bulk are clean enough that they are both fully
gapped.  Thus only Andreev scattering along the edge contributes to
$\si_\text{SC-Hall}$.  To include the effects of charge conserving inelastic
scattering, we first divide the IQH edge channel into many segments each of
length $l_\phi$ -- the dephasing length.  Each segment is coupled to a
superconductor (see Fig.  \ref{SCedge}a) which induces the coherent Andreev
scattering: free electrons up to a chemical potential $\mu$ can be coherently
scattered and come out as holes.  The incoming edge state is an equilibrium
state with an incoming chemical potential $\mu$, while the outgoing edge state
out of one SC segment is not an equilibrium state.  Charge conserving inelastic
scattering equilibrates the outgoing edge state, which now has an chemical
potential $\mu'$.  From $\mu-\mu'$, we can determine $\sigma_{\text{SC-Hall}}$
for the segment.

To analyze the change in $\mu$ after passing a single SC segment,
let us start with the equation of motion for free chiral fermion:
\begin{align}
\begin{split}
\ii\hbar \dot c &=  v_f (-\ii \partial_x-k_F) c
 +\frac {\ii\hbar}{2} [\vsc\partial_x c^\dagger +\partial_x (\vsc c^\dagger)],
\\
\ii\hbar \dot c^\dagger &=v_f (-\ii \partial_x+k_F)  c^\dagger 
+\frac {\ii\hbar}{2}  [\vsc^* \partial_x c +\partial_x (\vsc^*  c)],
\end{split}
\end{align}
where $v_f$ is the velocity of the chiral fermion, $k_F$ is Fermi momentum, and $\vsc(x)$ is the
SC coupling coefficient which depends on $x$ ($\vsc=0$ for edge
not under the superconductor).  
We treat $(c,c^\dag)=(\psi_1,\psi_2)\equiv \psi^T$ as independent fields.
For a mode with a frequency $\om$, the  equation of motion becomes
\begin{align}
 \om \psi = 
\bpm
v_f (-\ii \partial_x-k_F) & \frac \ii 2 (\vsc \partial_x +\partial_x \vsc ) \\
\frac \ii 2 (\vsc^* \partial_x +\partial_x \vsc^* ) & v_f (-\ii \partial_x+k_F)\\
\epm 
\psi
\end{align}
or (up to linear $\vsc$ order)
\begin{align}
& 
-v_f
\bpm
1  & \frac{\vsc}{2v_f} \\
\frac{\vsc^*}{2v_f} & 1 \\
\epm^{-1} 
  \ii \partial_x 
\bpm
1  & \frac{\vsc}{2v_f} \\
\frac{\vsc^*}{2v_f} & 1 \\
\epm^{-1}
\psi 
\nonumber\\
& \approx 
\bpm
\om + v_f k_F & 0\\
0 & \om - v_f k_F\\
\epm 
\psi .
\end{align}
Let $\t\psi = \bpm
1  & \frac{\vsc}{2v_f} \\
\frac{\vsc^*}{2v_f} & 1 \\
\epm^{-1}
\psi 
$, we can rewrite the above as
\begin{align}
&
- \ii \partial_x \t \psi(x) = M(x)\t \psi(x),
\\
M (x)&=
\bpm
1  & \frac{\vsc}{2v_f} \\
\frac{\vsc^*}{2v_f} & 1 \\
\epm
\bpm
\frac{\om}{v_f} + k_F & 0\\
0 & \frac{\om}{v_f} -  k_F\\
\epm 
\bpm
1  & \frac{\vsc}{2v_f} \\
\frac{\vsc^*}{2v_f} & 1 \\
\epm
\nonumber\\
& \approx \frac{\om}{v_f}
\bpm
1  & \frac{\vsc(x)}{v_f} \\
\frac{\vsc^*(x)}{v_f} & 1 \\
\epm
+
\bpm
k_F & 0 \\
0 &  -k_F \\
\epm
.
\nonumber 
\end{align}
Solving the above differential equation, we find
$\t\psi(x) =P[\ee^{\ii  \int_0^x \dd x M(x)}]  \t\psi(0)$,
where $P$ is the path ordering.
Now we assume that $\vsc(x)=0$ for $x<0$ and $x>l_\phi$, and 
$\vsc(x)$ is a constant for $x\in [0,l_\phi]$.
We find
$\psi(l_\phi) = S \psi(0)$, where the unitary matrix $S$ is given by
\begin{align*}
S = P[\ee^{\ii  \int_0^{l_\phi} \dd x M(x)}] 
=
\ee^{\ii\phi}
\begin{pmatrix}
\ee^{\ii k_Fl_\phi} \cos\th  &  \ii\ee^{\ii\vphi} \sin \th\\
\ii \ee^{-\ii\vphi} \sin \th & \ee^{-\ii k_Fl_\phi } \cos \th \\
\end{pmatrix}
\end{align*}
and,  to the linear order in $\vsc$, the scattering angle is
\begin{align}
\label{thom}
\th \approx \frac{|\vsc| \om }{k_F v_f^2}\sin (k_F l_\phi).
\end{align}


The modes with a frequency $\omega$ are electron-like state with momentum $k+k_F$ and hole-like state with momentum $-k+k_F$, where $k=\frac{\omega}{v_f}$.
Denote $a_k, b_k$ as incoming and outgoing electron annihilation operator of
momentum $k$ measured from $k_F$. $b_k$ is determined by 
$b_k = S_{11} a_k+S_{12} a_{-k}^\dagger$.

In the zero temperature limit, the occupation numbers of incoming and outgoing
electrons are $\langle a_k^\dagger a_k\rangle = 1$ for $k\leq
\frac{\mu}{\hbar v_f}$,  $\langle a_k^\dagger a_k\rangle = 0$ for $k >
\frac{\mu}{\hbar v_f}$, and
\begin{align}
\langle b_k^\dagger b_k\rangle =&\cos^2 \theta \langle a_k^\dagger a_k\rangle +\sin^2\theta  \left(1-\langle a_{-k}^\dagger a_{-k}\rangle\right)\nn\\
=&\begin{cases}
0 , &k>\frac{\mu}{\hbar v_f}\\
\cos^2 \left(\theta(k)\right), & -\frac{\mu}{\hbar v_f}\leq k\leq \frac{\mu}{\hbar v_f}\\
1, & k<-\frac{\mu}{\hbar v_f}
\end{cases}
\end{align}
The outgoing electrons relax to $\mu'$ with the same density
\begin{align}
\int_{-\frac{\mu}{\hbar v_f}}^{\frac{\mu}{\hbar v_f}}& \frac{dk}{2\pi} 
\cos^2 \left( \frac{|\vsc| \sin(k_F l_\phi)}{v_f k_F}  k\right)
 =\int_{-\frac{\mu}{\hbar v_f}}^{\frac{\mu'}{\hbar v_f}} \frac{dk}{2\pi}\\
\Rightarrow \quad &\mu'=\frac{\hbar v_f^2 k_F}{2 |\vsc| \sin(k_F l_\phi)}
\sin \frac{2|\vsc| \sin(k_F l_\phi)\mu}{\hbar v_f^2 k_F} \label{mun}
\end{align}
When $\frac{|\vsc| \mu}{\hbar v_f^2 k_F}\ll 1$, we have
\begin{align}
&
\mu'= \mu \left( 1-\frac{1}{6}
\left(\frac{2|\vsc| \sin( k_F l_\phi)\mu}{\hbar v_f^2 k_F}\right)^2\right).
\nonumber 
\end{align}
This change of $\mu$ through one segment of length $l_\phi$ allows us to obtain, for a length $\del L_\text{edge}$ edge,
\begin{align}
 \si_\text{SC-Hall} =-\frac{\delta \mu}{\mu} =
\Big( \frac{ \mu}{\Del} \Big)^2 \frac{\del L_\text{edge}}{l_\phi}
\label{sc-hall}
\end{align}
with $\frac{1}{\Del}= \sqrt{\frac{1}{3}}\frac{|\vsc|}{v_f^2 \hbar k_F }$, where
we have replaced $\sin^2(k_F l_\phi)$ by its average $\frac 12$.  Interestingly, $\si_\text{SC-Hall}$ is
proportional to $\mu^2$, or rather, non-Ohmic.

In the high temperature limit, 
\begin{align}
\begin{split}
&\langle a_k^\dagger a_k\rangle = g(\mu,k)\equiv \frac{1}{e^{\frac{\hbar v_f k-\mu}{k_BT}}+1}\\
\langle b_k^\dagger b_k\rangle =& \cos^2 \theta g(\mu,k)+\sin^2\theta (1-g(\mu,-k))\\
 =&\cos^2 \theta g(\mu,k)+\sin^2\theta g(-\mu,k).
\end{split}
\end{align}
Keep to the first order of $\frac{\mu}{k_BT}$ and $\frac{\vsc}{v_f}$, we reach
\begin{align}
 \mu' &=\mu \Big[1- \frac{2\pi^2}{3} 
\left(\frac{|\vsc| \sin(k_F l_\phi)}{ v_f k_F }\frac{k_BT}{\hbar v_f}\right)^2 \Big].
\end{align}
From this we obtain, for a length $\del L_\text{edge}$ edge,
\begin{align}
\label{SCHallT}
 \si_\text{SC-Hall} =  \ga \frac{\del L_\text{edge}}{l_\phi} .
\end{align}
with $\gamma=\frac{\pi^2}{3} \left(\frac{|\vsc| k_BT}{\hbar v_f^2 k_F
}\right)^2$.
In this case, $\si_\text{SC-Hall}$ is independent of $\mu$ and is Ohmic.

If either the SC film or IQH bulk are not clean enough and have gapless
electronic states that couple to the chiral edge channel, we can take into
account those gapless states by assuming the superconductor to be a gapless
superconductor.  In this case, $\si_\text{SC-Hall}$ will in addition receive a
contribution from the electron tunneling into the quasiparticle states in the
gapless superconductor.  We expect such a contribution to be Ohmic and
$\si_\text{SC-Hall}$ can be modeled by \eqref{SCHallT} in all temperature
range.

In the following, we will separately calculate $\si_{12}(B)$, using the  non-Ohmic (\ref{sc-hall}) or Ohmic (\ref{SCHallT})
$\si_\text{SC-Hall}$.

\begin{figure}[tb] 
\centering 
\includegraphics[height=1.5in]{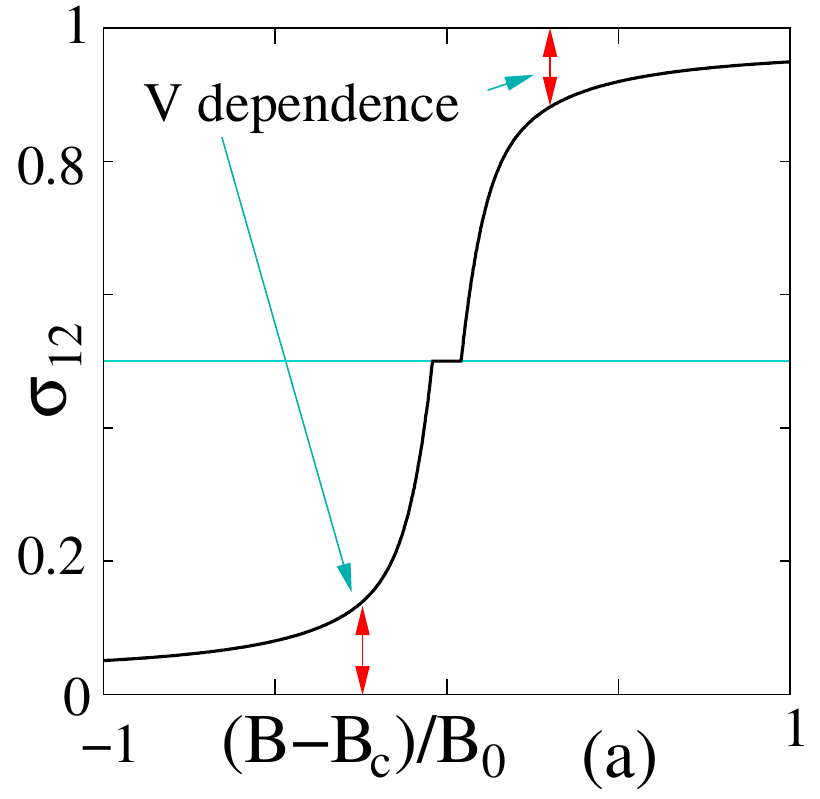} 
\includegraphics[height=1.5in]{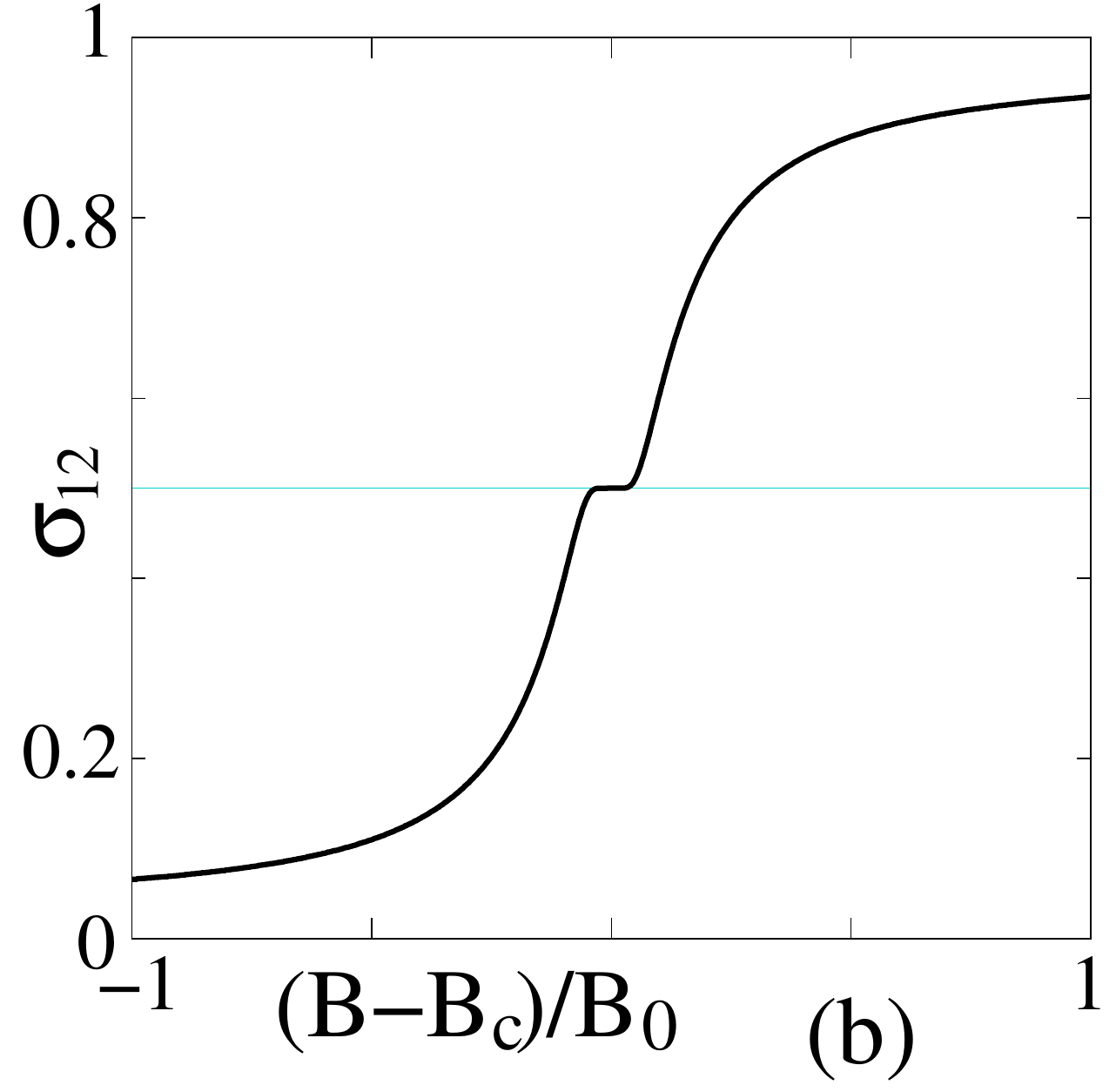} 
\caption{
Two terminal conductance $\si_{12}$ as a function of magnetic field $B$.  (a)
Non-Ohmic case \eqref{si12B}, with $l_\text{edge}/a=70$ and $\frac{2 \mu_0^2
l_\text{edge}}{\Del^2 l_\phi}  = 0.12$.  Deviation of $\si_{12}$ from
$\frac{e^2}{h}$ and $0$ will have a clear voltage $V=\mu_0/e$ dependence.  (b)
Ohmic case \eqref{si12BT}, with $\ga \frac{l_\text{edge}}{l_\phi}=1/14$.  The
curve for Ohmic case is independent of  the percolation cut-off length scale
$a$.
}
\label{si12} 
\end{figure}

\noindent \textbf{Non-Ohmic case}: 
From \eqref{sc-hall}, we see that the contact resistance  can be much bigger
than $\frac{h}{e^2}$, as long as $\mu^2\del L_\text{edge}$ is small enough.
The current $\del I=\si_\text{SC-Hall}\ \mu$ flowing from the edge to the superconductor
will cause a drop in the chemical potential $\mu$ along the edge:
\begin{align}
\dd \mu(x) =  
-\si_\text{SC-Hall}\ \mu =
- \frac{\mu^3(x) }{l_\phi \Del^2}
\dd x
\end{align}
Solving the above equation, we find $ \mu = \mu(L_\text{edge})=
{\mu_0}/{\sqrt{ \frac{2 \mu_0^2}{\Del^2 l_\phi} L_\text{edge}+ 1}}$ for an edge
of length $L_\text{edge}$. 

Therefore, for $B>B_c$ (see Fig. \ref{HallSC}a)
\begin{align}
 \si_{12} &= \frac{\mu_0+\mu}{2\mu_0}
=\frac{\mu_0+
\frac{\mu_0}{ \sqrt{ \frac{2 \mu_0^2}{\Del^2 l_\phi} L^>_\text{edge}+ 1 } }
}{2\mu_0},
\end{align}
In a percolation cluster of size $\xi$, the edge length is $\frac{\xi^2}{a}$,
where $a$ is the cut-off length scale of the percolation model. The  total edge
length is $ L^>_\text{edge} = \frac{l_\text{edge}}{\xi} \frac{\xi^2}{a}
=l_\text{edge}\frac{\xi}{a}$.  The linear size of the percolation cluster $\xi$
scales as 
\begin{align}
\label{xi}
 \xi = a \Big( \frac{|B_c-B|}{B_0} \Big)^{-\nu} +a,\ \ \ \nu=1.33
\end{align}
With the above choice, we see that $(L^>_\text{edge},\si_{12}) \to
(l_\text{edge},1)$ as $B\to \infty$ (assuming $\frac{2 l_\phi \mu_0^2}{v^2 \hbar^2} l_\text{edge}$ is small), and
$(L^>_\text{edge},\si_{12}) \to (\infty, \frac12) $ as $B\to B_c$.

But $\xi$ can only increase up to $l_\text{edge}$, the width of superconductor
covered Hall bar, beyond which $\xi$ remains to be $l_\text{edge}$ in the
metallic phase in Fig. \ref{HallSC}b .  To model such a behavior, we choose
\begin{align}
 L^>_\text{edge} &=a^{-1}\xi l_\text{edge} \Th(B-B_c)\Th(l_\text{edge}-\xi) 
\nonumber\\
& 
+ a^{-1}l_\text{edge}^2\Th(\xi-l_\text{edge})
\\
& 
+ a^{-1}l_\text{edge}^2\ee^{(l_\text{edge}-\xi)/\xi} \Th(B_c-B)\Th(l_\text{edge}-\xi).
\nonumber 
\end{align}
where $\Th(x)=1$ if $x>0$ and $\Th(x)=0$ if $x<0$.  When $B>B_c$, the above
gives $ L^>_\text{edge}=a^{-1}\xi l_\text{edge}$ or $a^{-1} l_\text{edge}^2$
near $B_c$ (see Fig. \ref{SCedge}b).  When $B$ is much less than $B_c$, we also
assign $L^>_\text{edge}$ a very large value to make $\mu_0/\sqrt{
\frac{2 \mu_0^2}{\Del^2 l_\phi} L_\text{edge}^>+ 1 } $ vanishes.  This allows
us to combine the $B>B_c$ and $B<B_c$ results together later.  For $B<B_c$ (see
Fig. \ref{HallSC}c)
\begin{align}
 \si_{12} &= \frac{\mu_0-\mu}{2\mu_0}
=\frac{\mu_0-
\frac{\mu_0}{\sqrt{ \frac{2 \mu_0^2}{\Del^2 l_\phi} L^<_\text{edge}+ 1 } }
}{2\mu_0} 
\end{align}
where
\begin{align}
 L^<_\text{edge} &=a^{-1}\xi l_\text{edge} \Th(B_c-B)\Th(l_\text{edge}-\xi) 
\nonumber\\
& 
+ a^{-1}l_\text{edge}^2\Th(\xi-l_\text{edge})
\\
& 
+ a^{-1}l_\text{edge}^2\ee^{(l_\text{edge}-\xi)/\xi} \Th(B-B_c)\Th(l_\text{edge}-\xi)
\nonumber 
\end{align}

We can combine the $B>B_c$ and $B<B_c$ cases: 
\begin{align}
\label{si12B}
 \si_\text{12}
=\frac{1}{2}\left(1
+\frac{1}{\sqrt{ \frac{2 \mu_0^2}{\Del^2 l_\phi} L^>_\text{edge}+ 1 } }
-\frac{1}{\sqrt{ \frac{2 \mu_0^2}{\Del^2 l_\phi} L^<_\text{edge}+ 1 } }\right)
\end{align}
With the above design of $L^>_\text{edge}$ and $L^<_\text{edge}$, only one of
the two terms in $\frac{1}{\sqrt{\frac{2 \mu_0^2}{\Del^2 l_\phi}
L^>_\text{edge}+ 1 } } -\frac{1}{\sqrt{\frac{2 \mu_0^2}{\Del^2 l_\phi}
L^<_\text{edge}+ 1 } }$ contributes in either the $N_\text{Chern}=1$ phase or
the $N_\text{Chern}=0$ phase.  In the metallic phase (see Fig. \ref{HallSC}b),
both terms are small, and their difference makes the contribution even smaller.
This gives rise to $\frac12$ quantized two terminal conductance.  The above
result is plotted in Fig. \ref{si12}a.  Such a result is very close to what was
observed in \Ref{HW160605712}.  But it has a very different mechanism than what
was proposed in \Ref{CZ10082003,WZ150700788}.  In our non-Ohmic case, the
$\si_{12}=\frac12 \frac{e^2}{h}$ plateau roughly corresponds to the metallic
phase in Fig.  \ref{HallSC} where $\xi / l_\text{edge}\approx 1$, with no need
to introduce 1D chiral Majorana fermion on the edge.

\noindent \textbf{Ohmic case}: 
From \eqref{SCHallT}, we see that the contact resistance  can be much bigger
than $\frac{h}{e^2}$, if $\ga \del L_\text{edge}/l_\phi$ is small enough.  From
the equation $ \dd \mu(x) =  -\ga \frac{\dd x }{l_\phi } \mu(x) $ and for a
given total length of the edge channel $L_\text{edge}$, we find $ \mu = \mu_0
\ee^{ - \ga L_\text{edge}/l_\phi} $.  Therefore, for $B>B_c$ (see Fig.
\ref{HallSC}a) $ \si_{12} = \frac{\mu_0+\mu}{2\mu_0} =\frac{1+ \ee^{ - \ga
L_\text{edge}/l_\phi} }{2}$, where $ L_\text{edge} = \frac{l_\text{edge}}{\xi}
\frac{\xi^2}{a} =l_\text{edge}\frac{\xi}{a}$.  With $\xi$ given in \eqref{xi},
we see that $L_\text{edge} \to l_\text{edge}$ as $B\to \infty$ and
$L_\text{edge} \to \infty $ as $B\to B_c$.  Similarly, for $B<B_c$ (see Fig.
\ref{HallSC}c), $ \si_{12} = \frac{\mu_0-\mu}{2\mu_0} =\frac{1- \ee^{ - \ga
L_\text{edge}/l_\phi}}{2}$.  We can combine the $B>B_c$ and $B<B_c$ cases
together and obtain 
\begin{align}
\label{si12BT}
 \si_\text{12} 
&=\frac{1 + \text{sgn}(B-B_c) 
\ee^{-( \frac{B_0^{\nu}}{|B_c-B|^{\nu}}  +1 ) 
\frac{\ga l_\text{edge}}{l_\phi} }}{2} 
.  
\end{align}
The above result is plotted in Fig. \ref{si12}b.  Such a result for Ohmic case
is also very close to what was observed in \Ref{HW160605712}.  But for Ohmic
case, the $\si_{12}=\frac12 \frac{e^2}{h}$ plateau is much broader than the
metallic phase in Fig.  \ref{HallSC}.

\noindent
\textbf{Summary:} In the percolation model, 
we considered two possible cases, Ohmic case and non-Ohmic case, both can explain the $\sigma_{12}(B)$ cuve in the 
experiment \Ref{HW160605712}.  More experiments are needed to see which case
applies. If an Ohmic contact conductance is observed, it will indicate either
the SC and/or IQH bulks have gapless electronic states, or the electron
temperature is high.  

If a non-Ohmic contact conductance $\si_\text{SC-Hall}$ between the
superconductor and the IQH edge channel is observed near $\sigma_{12} \sim 0$
or $\sigma_{12} \sim 1$, it will indicate the SC and IQH bulks to be fully
gapped. Therefore observing a non-Ohmic contact conductance is a sign of clean
samples, which is necessary for further strong quantum coherent phenomena. For
instance, on such samples at low enough temperature, the dephasing length can
become large, and 1D chiral Majorana fermions can appear.

After posting this paper, another paper \Ref{HS170806752} was posted where the
same conclusion was reached via a similar consideration.  A month later, yet
another paper \Ref{LZ170905558} was posted, where the dephasing length $l_\phi$
is assumed to be larger than the ``$p+\ii p$ SC coherence length'' $\xi_{p+\ii
p}$ (put it another way, the minimum width of a $p+\ii p$ SC stripe such that
1D chiral Majorana fermions on the two edges are well separated). In this case,
the 1D chiral Majorana edge mode can be well defined, and give rise to a
$\frac12 \frac{e^2}{h}$ plateau in $\si_{12}$.  In this paper, we consider the
opposite limit $l_\phi < \xi_{p+\ii p}$ without coherent 1D chiral Majorana
edge mode, and show that there is still a $\frac12 \frac{e^2}{h}$ plateau.
Furthermore, the $B$ dependence of $\si_{12}$ can agree with the experiment
very well, with a proper choice of some parameters.  In particular, if we
choose $B_0 \sim 200$mT, the plateau width will be about $20$mT (see Fig.
\ref{si12}).

We would like to thank K. L. Wang, Yayu Wang, and Shoucheng Zhang for very
helpful discussions.  This research was supported by NSF Grant No.  DMR-1506475
and NSFC 11274192.

\appendix

\section{A general property of scattering matrix $S$}

Denote $a_k, b_k$ as incoming and outgoing electron annihilation operator of
momentum $k$ on the IQH edge (see Fig. \ref{SCedge}a).  After passing under the
superconductor and an energy conserving Andreev scattering, $b_k$ is determined
by a scattering matrix $S$:
\begin{align}
\psi^\text{out}_k & = S(k) \psi^\text{in}_k,
\nonumber\\
\psi^\text{out}_k & \equiv
\bpm
b_k\\
b^\dagger_{-k}
\epm
,\ \ \ \ \
\psi^\text{in}_k \equiv
\bpm
a_k\\
a_{-k}^\dagger
\epm
\end{align}
Since $S(k)$ must preserve the anti-commutation relation
before and after the Andreev scattering:
\begin{align}
 \{\psi_{k,i}^\dag, \psi_{k,j}\}=\del_{ij}, \ \ \
 \{\psi_{k,i}, \psi_{-k,j}\}=\si^x_{ij},
\end{align}
thus $S(k)$ must satisfy
\begin{gather}
 S^\dag(k)S(k)=1 
,
\nonumber\\
 S(k)\si^x S^\top(-k)=\si^x
\ \ \text{ or } \ \ 
\si^x S(k)\si^x = S^*(-k)
.
\end{gather}
A general form of $U(2)$ matrix is parametrized as
\begin{align}
 S(k)=e^{\ii \varphi /2}\begin{pmatrix}
e^{\ii \phi_1} \cos \theta  &  e^{\ii \phi_2} \sin \theta \\
-e^{-i\phi_2} \sin \theta  &  e^{-i\phi_1} \cos \theta
\end{pmatrix}
\label{sm}
\end{align}
where $\theta, \phi_1, \phi_2, \varphi$ all depends on $k$.
At $ k=0$, we obtain the restriction on $\theta,  \phi_1, \phi_2,\varphi$ 
\begin{align}
\begin{split}
&\ \ \ \
e^{\ii \varphi/2}\begin{pmatrix}
e^{\ii \phi_1} \cos \theta  &  e^{\ii \phi_2} \sin \theta \\
-e^{-i\phi_2} \sin \theta  &  e^{-i\phi_1} \cos \theta
\end{pmatrix}
\\
&=e^{-i\varphi/2}\begin{pmatrix}
e^{\ii \phi_1} \cos \theta  &  -e^{\ii \phi_2} \sin \theta \\
e^{-i\phi_2} \sin \theta  &  e^{-i\phi_1} \cos \theta
\end{pmatrix}
\end{split}
\label{s0}
\end{align}
Eq (\ref{s0}) has two sets of solutions,
\begin{enumerate}
\item $\cos \theta \neq 0, \sin\theta=0$, therefore
\begin{align}
\theta=&\pi n,\; \varphi=2\pi m, \nn\\
 S_0=&\pm  \begin{pmatrix}
e^{\ii \phi_1} & 0 \\
0  & e^{-i\phi_1} 
\end{pmatrix}
\end{align}
That is, $b_0=\pm e^{\ii \phi_1}a_0,\  b_0^\dagger b_0=a_0^\dagger a_0$, a pure
transmission except a phase shift.  \item $\cos\theta=0,\sin\theta\neq 0$,
therefore
\begin{align}
\theta=&\frac{\pi}{2}+\pi n,\; \varphi=(2m+1)\pi,\nn\\
 S_0=&\pm e^{\ii \frac{\pi}{2}} \begin{pmatrix}
0 & e^{\ii \phi_2} \\
-e^{-i\phi_2}  & 0
\end{pmatrix}
\end{align}
That is, $b_0=\pm \ii e^{\ii \phi_2}a_0^\dagger,\ b_0^\dagger b_0=a_0
a_0^\dagger $, a pure Andreev transmission.  \end{enumerate} So for weak SC
coupling $\vsc$, $S_{12}(k) \to 0$ as $k\propto \om \to 0$, agreeing with our
explicit calculation \eqref{thom}.

\bibliography{../../bib/wencross,../../bib/all,../../bib/publst}

\end{document}